\documentclass[aps,pra,twocolumn,showpacs]{revtex4-1}

\usepackage{amsmath} 
\usepackage{graphicx}

\begin{document}

\title{Strategies for choosing path-entangled number states for optimal robust quantum optical metrology in the presence of loss}

\author{Kebei Jiang}
\email{kjiang2@tigers.lsu.edu}
\author{Chase J.\ Brignac}
\author{Yi Weng}
\author{Moochan B.\ Kim}
\author{Hwang Lee}
\author{Jonathan P.\ Dowling}
 \affiliation{Hearne Institute for Theoretical Physics and Department of Physics and Astronomy \\
Louisiana State University, Baton Rouge, LA 70803 }

\date{\today}

\begin{abstract}

To acquire the best path-entangled photon Fock states for robust quantum optical metrology with parity detection, we calculate phase information from a lossy interferometer by using twin entangled Fock states. We show that $(\text{a})$ when loss is less than 50\% twin entangled Fock states with large photon number difference give higher visibility while when loss is higher than 50\% the ones with less photon number difference give higher visibility; $(\text{b})$ twin entangled Fock states with large photon number difference give sub-shot-noise limit sensitivity for phase detection in a lossy environment. This result provides a reference on what particular path-entangled Fock states are useful for real world metrology applications.

\end{abstract}

\maketitle


\section{Introduction}
\label{sec:intro}

The application of quantum states of light has long been proposed to achieve greater precision, resolution, and sensitivity than what is possible classically \cite{Dowling2008,PhysRevD.26.1817}. A maximally path-entangled state is a superposition of all photons in one path with none in the other, and vice versa. These states are known as $N00N$ states and were introduced to achieve high resolution and high sensitivity in metrology and imaging \cite{PhysRevLett.85.2733,PhysRevLett.99.070801}. They are defined as $\vert N::0\rangle_{a,b}=1/\sqrt{2}(\vert N,0\rangle_{a,b}+\vert 0,N\rangle_{a,b})$, where $a$ and $b$ indicate the two paths of a two-mode interferometer. However, $N00N$ states tend to decohere easily when photons are lost from the system. This makes $N00N$ states unusable in real life, where loss is almost always present \cite{PhysRevA.75.053805,Gilbert2007,RevSciInstrum.76.043103}. In 2008, Huver $\textit{et al.}$~proposed a class of generalized Fock states where decoy photons are introduced to the $N00N$ state in both arms of a two-mode interferometer \cite{PhysRevA.78.063828}. These are called $mm'$ states and they are denoted $\vert m::m'\rangle_{a,b}=1/\sqrt{2}(\vert m,m'\rangle_{a,b}+\vert m',m\rangle_{a,b})$. It was discovered that $mm'$ states have better metrological performance over $N00N$ states in the presence of photon loss. In this paper, we locate the best performing $m$ and $m'$ under certain fixed loss, where the photon number difference ($\Delta m=m-m'$) between the two arms in the initial state is fixed.

In addition to state preparation, achieving super-resolution (beating the Rayleigh limit \cite{ReschPRL2007,GiovannettiPRA2009}) and super-sensitivity (beating the shot-noise limit \cite{NagataSci2007}) requires detection schemes with particular properties. In this paper we choose parity detection, which reaches Heisenberg-limited sensitivity when combined with lossless $N00N$ states \cite{PhysRevA.61.043811,PhysRevA.68.023810,doi:10.1080/00107514.2010.509995}. The parity operator can be expressed as $\hat{\Pi}=\exp(i \pi \hat{n})$ in the number basis or $\hat{\Pi}=\exp(i \pi (\hat{J}_0-\hat{J}_{z}))$ in Schwinger notation \cite{PhysRevA.33.4033,PhysRevA.57.4004,doi:10.1080/09500340.2011.585251}, which will later be discussed in more detail. The parity operator is assigned a parity of $+1$ if the measured number of photon is even and a parity of $-1$ if odd. In this paper we use the parity operator transformed through a beam splitter. Readers who are interested in more details about the parity operator and its application in quantum optical metrology may refer to Ref. \cite{doi:10.1080/00107514.2010.509995}.

In this paper we calculate both the visibility and sensitivity of the phase signal from the interferometer. On one hand, the signal from parity detection can be negative therefore the ordinary definition of visibility is not applicable. We define a ``relative visibility'' in Section $\mathrm{V}$ to solve this problem. On the other hand, sensitivity is defined by the linear error propagation method as \cite{Bevington1969}
\begin {align}
\delta \phi=\frac{\Delta \hat{\Pi}}{\vert \partial \langle \hat{\Pi} \rangle /\partial \phi \vert},
\label{eq:DefSensitivity}
\end {align}
and $\Delta \hat{\Pi}=\sqrt{\langle \hat{\Pi}^{2} \rangle-\langle \hat{\Pi} \rangle ^{2}}$. A promising goal of this work is to provide a strategy for choosing the path-entangled number state that optimizes either visibility or sensitivity for a given loss.

\begin{figure}
	\centering
		 \includegraphics[width=0.4\textwidth]{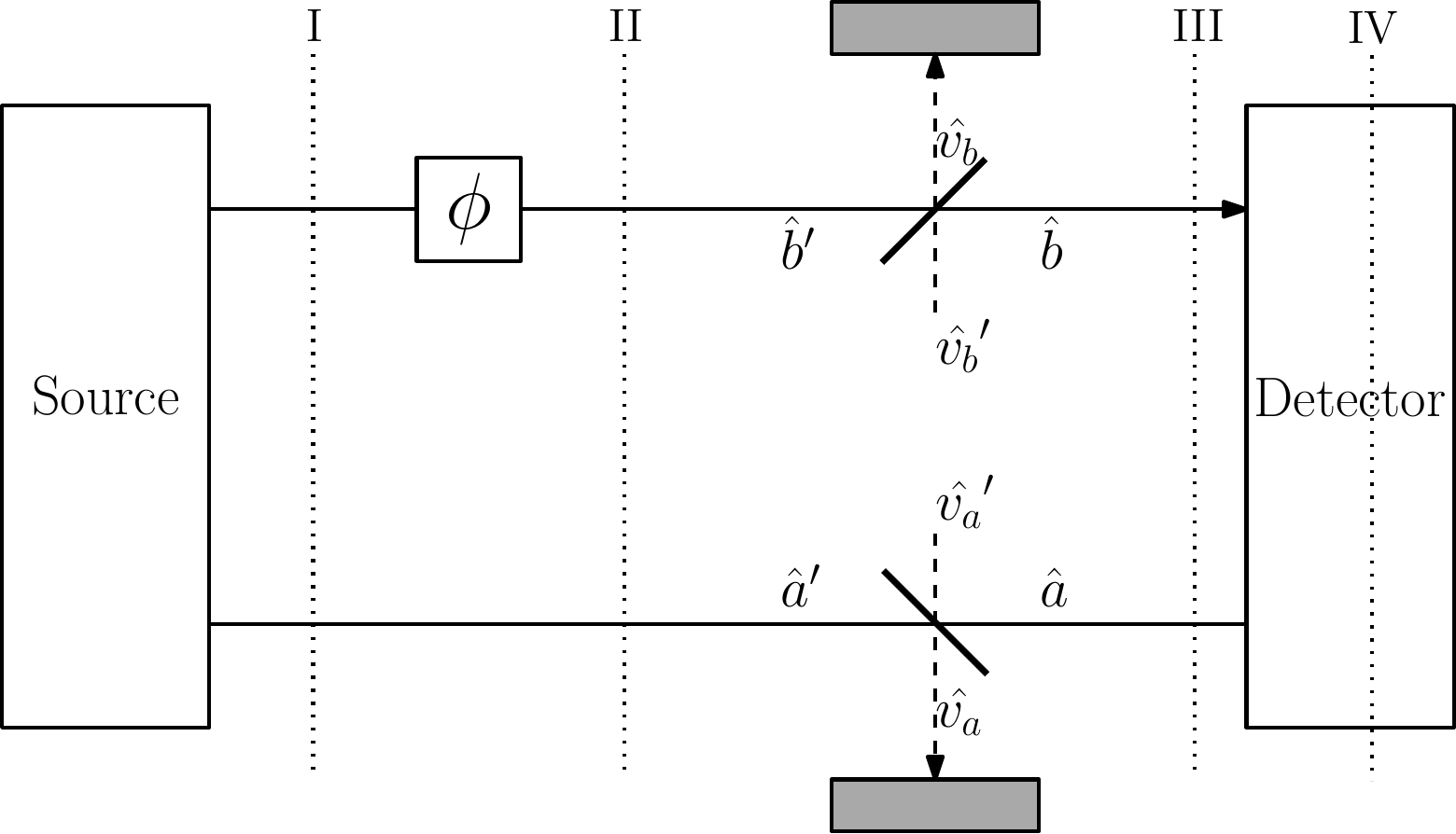}
	\caption{A lossy interferometer model which can be used to measure the accumulated phase on path $b$. The phase shift operator is given by $\hat{U}=\exp(i \phi \hat{b}^{\dagger}\hat{b})$. }
	\label{fig:interferometer}
\end{figure}

\section{Density Matrix}
We start with the classical Mach-Zehnder interferometer as shown in Fig.~\ref{fig:interferometer}, where the source and the detector are represented by their respective boxes. Similar to the approach in Ref. \cite{Loudon2000,PhysRevA.75.053805,Gardiner2004}, the loss in the interferometer is modeled by adding fictitious beam splitters. Notice that it will not change the density matrix if the beam splitter is placed before the phase shifter. This can be easily demonstrated by removing phase dependence from Eq.~\eqref{eq:DM} (see below) and applying $\hat{U}=\exp(i \phi \hat{b}^{\dagger}\hat{b})$ on it afterwards.

The wave function for the $mm'$ input state at stage I is
\begin{align} %
\vert m::m' \rangle_{a',b'} = \frac{1}{\sqrt{2}} ( \vert m, m' \rangle_{a',b'} + \vert m', m \rangle_{a',b'}). %
\end{align} %
Without loss of generality, we assume $\Delta m=m-m'$ is positive (a $mm'$ state reduces to a $N00N$ state when $m=N$ and $m'=0$).

Then the phase shifter introduces a phase shift $\phi$ on arm $b$ so that the state at stage II becomes (see Fig.~\ref{fig:interferometer})
\begin{align} %
\vert \Psi \rangle_{\mathrm{II}} & = \frac{1}{\sqrt{2}} \left( e^{im'\phi} \vert m, m' \rangle_{a',b'} + e^{im\phi} \vert m', m \rangle_{a',b'} \right)\\
\nonumber
& = \alpha | m, m' \rangle_{a',b'} + \beta | m', m \rangle_{a',b'}, %
\end{align} %
where $\alpha =e^{im'\phi}/\sqrt{2}$ and $\beta =e^{im\phi}/\sqrt{2}$. We can see that because of the different number of photons being phase-shifted on arm $b$, the two paths accumulated different phase shifts and thus provide the possibility of interference upon detection.

The mode transformation by the beam splitter is given by Ref.~ \cite{Gerry2005}%
\begin{align} %
\hat{a} = t_a^* \hat{a}' + r_a^* \hat{v}_a', \\ %
\nonumber
\hat{b} = t_b^* \hat{b}' + r_b^* \hat{v}_b', %
\end{align} %
where $t_i = \sqrt{T_i} \exp (i\varphi_i)$ and $r_i = \sqrt{R_i} \exp (i\psi_i)$ ($i=a,b$) are the complex transmission and reflectance coefficients for modes $a$ and $b$, where $T_i+R_i=1$.


By tracing out the vacuum modes on both paths, we have a density matrix $\rho_{ab}$ that corresponds to the output field as
\begin{align} %
\nonumber
\rho_{ab}= &~ \mathrm{Tr}_{v_a,v_b}[\vert \Psi \rangle_{\mathrm{III}} ~_{\mathrm{III}}\langle \Psi \vert]\\
\nonumber
 =&~\sum_{k=0}^{m} \sum_{k'=0}^{m'} (\vert \alpha \vert^2 d_1 \vert k,k'\rangle_{a,b} ~_{a,b}\langle k,k' \vert \\
\nonumber
& ~+\vert \beta \vert^2 d_2 \vert k',k\rangle_{a,b} ~_{a,b}\langle k',k \vert \\
\nonumber
& ~+\sum_{k=0}^{m'} \sum_{k'=0}^{m'} (\alpha \beta^* d_3 \vert \Delta m +k,k'\rangle_{a,b} ~_{a,b}\langle k,\Delta m +k' \vert\\
& ~+\alpha^* \beta d_4 \vert k',\Delta m +k\rangle_{a,b} ~_{a,b}\langle \Delta m +k',k \vert),
\label{eq:DM}
\end{align}
where coefficients $d_i (i={1,2,3,4})$ are defined as
\begin{align} %
\nonumber
d_1(k,k')=&~ \frac{1}{2} \binom{m}{k} \binom{m'}{k'} \vert T_a \vert^{k} \vert R_a \vert^{m-k} \vert T_b \vert^{k'} \vert R_b \vert^{m'-k'},\\
\nonumber
d_2(k,k')=&~ \frac{1}{2} \binom{m}{k} \binom{m'}{k'} \vert T_a \vert^{k'} \vert R_a \vert^{m'-k'} \vert T_b \vert^{k} \vert R_b \vert^{m-k}, \\
\nonumber
d_3(k,k')=&~  \frac{1}{2} \binom{m}{\Delta m +k}^{\frac{1}{2}} \binom{m}{\Delta m +k'}^{\frac{1}{2}} \binom{m'}{k}^{\frac{1}{2}} \binom{m'}{k'}^{\frac{1}{2}}, \\
\nonumber
     &\times T_a^{\frac{1}{2}(\Delta m+2k)} R_a^{m'-k} T_b^{\frac{1}{2}(\Delta m+2k')} R_b^{m'-k'} \\
\nonumber
d_4(k,k')=&~  \frac{1}{2} \binom{m}{\Delta m +k}^{\frac{1}{2}} \binom{m}{\Delta m +k'}^{\frac{1}{2}} \binom{m'}{k}^{\frac{1}{2}} \binom{m'}{k'}^{\frac{1}{2}} \\
      &\times T_a^{\frac{1}{2}(\Delta m+2k')} R_a^{m'-k'} T_b^{\frac{1}{2}(\Delta m+2k)} R_b^{m'-k}.
      \label{eq:DM coefficients}
\end{align} %
An equivalent way to describe the loss process is by using the Kraus operators and one may refer to Ref.~\cite{LeePRA2009,PhysRevLett.102.040403,PhysRevA.80.013825} and references therein.

\section{Parity Operator}

The calculation of this section is done in Schwinger notation, so we wish to discuss this representation briefly. Typical four-port two-mode interferometers can be described using Schwinger notation isomorphic to angular momentum \cite{PhysRevA.33.4033}. The operators are: $\hat{J}_{x}=(\hat{a}^{\dagger} \hat{b}+\hat{b}^{\dagger} \hat{a})/2, \hat{J}_{y}=(\hat{a}^{\dagger} \hat{b}-\hat{b}^{\dagger} \hat{a})/(2 i), \hat{J}_{z}=(\hat{a}^{\dagger} \hat{a}-\hat{b}^{\dagger} \hat{b})/2, \hat{J}_{0}=(\hat{a}^{\dagger} \hat{a}+\hat{b}^{\dagger} \hat{b})/2$ and $\hat{J}^2=\hat{J}_{x}^2+\hat{J}_{y}^2+\hat{J}_{z}^2$, which obey $[\hat{J}_{i},\hat{J}_{j}]=i \varepsilon_{ijk} \hat{J}_{k}$ and $[\hat{J}_{0},\hat{J}_{i}]=0 ~(i=x,y,z)$. Therefore the common eigenstate of $\hat{J}_{0}$ and $\hat{J}_{z}$ is the two-mode Fock state
\begin{align}
\vert j,\mu \rangle _z=\vert j+ \mu, j- \mu \rangle_{a,b}
\end{align}
with eigenvalues $j=\langle \hat{J}_0\rangle$ and $\mu=\langle \hat{J}_z \rangle $. One uses $\hat{J}^2$ in quantum mechanical angular momentum treatment where $\hat{J}_{0}$ is not well defined \cite{Sakurai1993}. However, $\hat{J}_{0}$ is useful in quantum optics because it is directly related to the total number of photons in the system.

In this paper we start with $\hat{\Pi}=(-1)^{\hat{J}_{0}-\hat{J}_{z}}$ at stage IV and transform it back to stage III as $\hat{\mathrm{Q}}$. The generator for the beam splitter transformation is $\hat{J_{x}}$, and we have
\begin{align}
\nonumber
\hat{\mathrm{Q}}&=\exp(-i \frac{\pi}{2} \hat{J}_{x})\hat{\Pi}\exp(i \frac{\pi}{2} \hat{J}_{x})\\
&=\exp(i \pi \hat{J}_{0})\exp(i \pi \hat{J}_{y}). %
\end{align}

Following Ref. \cite{doi:10.1080/09500340.2011.585251}, the parity operator inside the interferometer in number basis becomes
\begin{align} %
\hat{\mathrm{Q}}=\displaystyle\sum_{n=0}^{N}i^n \displaystyle\sum_{k=0}^{n}(-1)^k \vert k, n-k \rangle \langle n-k,k \vert,
\end{align} %
where the first summation is over all possible photon loss, and $N$ is the total number of photons without loss. It is easy to check that $\hat{\mathrm{Q}}^2=1$.

\section{Application of parity detection with loss and $mm'$ states}
With both the density matrix and the parity operator obtained at stage III, it is straightforward to calculate the expectation value of the parity operator for an $mm'$ state as
\begin{align}
\nonumber
\langle\hat{\mathrm{Q}}\rangle&=\mathrm{Tr}(\hat{\mathrm{Q}} \hat{\rho})\\
&=K_1+K_2 \cos\Delta m \phi,
\label{eq:expectation}
\end{align}

where $K_1$ and $K_2$ are defined as

\begin{align} %
\nonumber
K_1=& \displaystyle\sum_{k=0}^{m'}(d_1(k,k)+d_2(k,k)) \\
\nonumber
=& \left ( R_a^{m'} R_b^{m}+ R_a^{m} R_b^{m'} \right ) \\
   & \times {}_2 F_1(-m,-m';1;\frac{T_a T_b}{R_a R_b}), \\
\nonumber
K_2=& \displaystyle\sum_{k=0}^{m'}(d_3(k,k)+d_4(k,k)) \\
\nonumber
=& R_a^{m'} R_b^{m'} {T_a}^{\frac{\Delta m}{2}} {T_b}^{\frac{\Delta m}{2}} \binom{m}{\Delta m} \\
   & \times {}_2 F_1(-m',-m';1+\Delta m;\frac{T_a T_b}{R_a R_b}).
\label{eq:signal}
\end{align}
Here ${}_2 F_1(a,b;c;z)=\displaystyle\sum_{n=0}^{\infty}\frac{(a)_n (b)_n}{(c)_n}\frac{z^n}{n!}$ is the ordinary hypergeometric function \cite{Gradshteyn1965}. The Pochhammer symbol within are defined to be
\begin{align}
\nonumber
   (x)_n=\begin{cases}                          
    1, & \text{if $n=0$};\\
    x(x+1)\dots(x+n-1), & \text{if $n>0$},
  \end{cases}
\end{align}
which truncates the infinite summation in the hypergeometric function at $n=m'$.
Note Eq. \eqref{eq:expectation} and \eqref{eq:signal} reduce to the $N00N$ state result if $m=N$ and $m'=0$.

For later calculations and analysis in this paper we use loss rate $L_i \equiv 1-T_i ~ (i=a,b)$ instead of the transmission rate $T_i$ following traditional notation in metrology.

\section{Visibility}

We use the parity operator for detection and its expectation value can be negative in certain regions of parameter space. To quantify the degree of measured phase information we need a proper definition of visibility. From Eq.~\eqref{eq:signal} we can see that $K_1$ decreases and $K_2$ increases as the loss rate decreases. Hence, $K_1$ and $K_2$ have a range from 0 to 1 and so it is reasonable to define the measured signal as
\begin{align}
 S=\frac{K_2}{K_1+K_2}
\end{align}
which is always positive. We can then define a visibility related to the highest phase information degree (i.e. strongest signal) as \begin{align}
V(L_a,L_b)=\frac{S(L_a,L_b)}{S(0,0)}
\label{eq:visibility}
\end{align}
where $S(0,0)$ represents the signal without loss. It is easy to see this relative visibility has a value ranging from 0 to 1.

\subsection{Visibility for general cases}

\begin{figure}
\centering
	\includegraphics[width=0.45\textwidth]{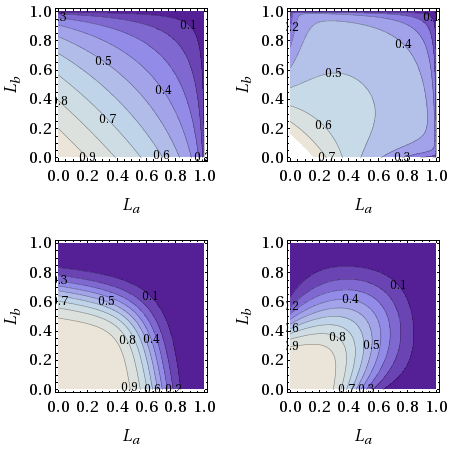}
	\caption{(Color online) From left to right, top to bottom. Visibility for $mm'$ states with $\Delta m=1$ ($\vert 1::0 \rangle$ and $\vert 3::2 \rangle$) and $mm'$ states with $\Delta m=4$ ($\vert 4::0 \rangle$ and $\vert 6::2 \rangle$), as a function of loss in both arms of two-mode interferometer, $L_a$ and $L_b$, respectively. Contour lines represent the value of the visibility.}
	\label{fig:visibility1}
\end{figure}

In Fig.~\ref{fig:visibility1}, we plot visibility as a function of loss rate. We see that, for $mm'$ states with a large total number of photons, the visibility changes rapidly at high or low loss but slowly at mild loss.

\begin{figure}
\centering
	\includegraphics[width=0.5\textwidth]{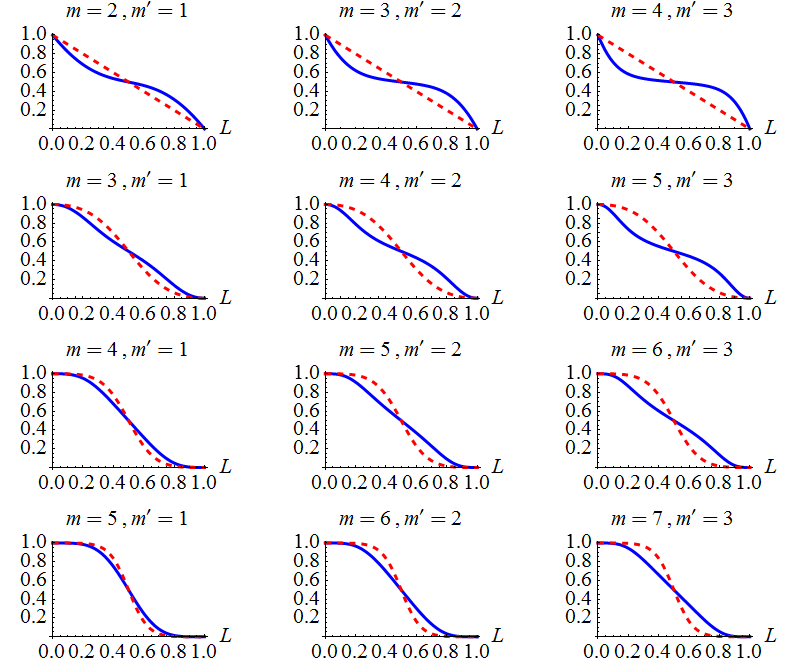}
	\caption{(Color online) Visibility for $mm'$ states as a function of loss $L$ in both arms of the two-mode interferometer. Each row has a fixed photon number difference and the total number of photons increases from left to right; each column has a fixed $m'$ and photon number difference increases from top to bottom. The red dashed lines represent the corresponding $N00N$ states with $N=\Delta m$.}
	\label{fig:visibility2}
\end{figure}

To clearly see the effect of photon number on visibility, we assume $L_a=L_b=L$ and plot visibility as a function of $L$ for different states in Fig.~\ref{fig:visibility2}. We observe $mm'$ states exhibit a lower visibility than corresponding $N00N$ states for loss rates lower than 50\%, and exhibit higher visibility for loss greater than 50\%. Each row has a fixed photon number difference and the total number of photons increases from left to right. We can see that with increasing total number of photons, the distance between the $mm'$ state and the $N00N$ state curves increases. Each column has fixed $m'$ and the photon number difference increases from top to bottom. We can see that as the photon number difference increases, the distance between the $mm'$ state and the $N00N$ state curves decreases.

Therefore, to obtain the best visibility, under photon loss less than 50\%, $N00N$ states should be used with as many photons as possible, i.e. the bottom right corner of the figure; for loss greater than 50\%, $mm'$ states should be used with as many photons as possible while keeping photon number difference to a minimum, i.e. the upper right corner of the figure.

Mathematically, the above results can be explained by expanding the visibility of any $mm'$ state around $L=\frac{1}{2}$ as
\begin{align}
V\vert _{L\approx\frac{1}{2}}=\frac{1}{2}+\frac{\Delta m^2}{m+m'} (L-\frac{1}{2})+O[(L-\frac{1}{2})^2].
\label{eq:vishalf}
\end{align}
Note that any $N00N$ state is just the corresponding $mm'$ state with $m'=0$, therefore it gives the steepest slope around $L=1/2$ in every case. Physically speaking, this result is different from Ref. \cite{PhysRevA.78.063828}, where $N00N$ states always have lower visibility than $mm'$ states under any loss. The reason for this discrepancy is that all off-diagonal terms of the density matrix are included  in Ref. \cite{PhysRevA.78.063828} while here the parity operator collects only part of the off-diagonal terms, making the amplitude of the resultant signal smaller.


\subsection{Visibility for extreme cases}
For situations with almost no loss, i.e. the loss rate $L\rightarrow 0$, the visibility function Eq.\eqref{eq:visibility} can be expanded as
\begin{align}
V\vert_{L\approx 0}=\binom{m}{\Delta m} L^{\Delta m}+ O[L^{\Delta m+1}],
\end{align}
which explains the behaviors of visibility curves around $L=0$ for different $\Delta m$ in Fig.~\ref{fig:visibility2}. Similarly, visibility for very lossy situations can be easily expanded as
\begin{align}
V\vert_{L\approx 1}=1-\binom{m}{\Delta m} (1-L)^{\Delta m}+ O[(1-L)^{\Delta m+1}]
\end{align}
because of the symmetry of the system. Another example for symmetry is that for $50\%$ loss the visibilities are calculated to be exactly one half for all $m$ and $m'$ value, which is the consequence of Eq.\eqref{eq:vishalf}.

\section{Sensitivity}

Another important quantity in quantum optical metrology is the precision, or sensitivity, of the phase measurement. The Heisenberg limit for any $mm'$ state under loss rate $L_a$ and $L_b$ should be $1/\widetilde{N}$ while the corresponding shot-noise limit is $1/\sqrt{\widetilde{N}}$ where $\widetilde{N}=(m+m')(1-L_a/2-L_b/2)$ is the effective number of transmitted photons. Therefore one usually compares the performance of different states with the same total number of photons. However, in order to keep the same resolving power we fix the photon number difference $\Delta m$ between two arms of the two-mode interferometer in this section.

\subsection{Sensitivity for general loss}

Sensitivity calculated from Eq. \eqref{eq:DefSensitivity}, \eqref{eq:expectation} and \eqref{eq:signal} can be expand as
\begin{align}
\delta \phi=\begin{cases}
 &\frac{1}{\Delta m}+\frac{(m+m')}{\Delta m}\csc (\Delta m \phi)L + O[L^2], \\ &\text{if $\Delta m~ \text{is even}$};\\
 &\frac{1}{\Delta m}+\frac{(m+m')}{\Delta m}\sec (\Delta m \phi)L + O[L^2], \\ &\text{if $\Delta m~ \text{is odd}$}.
 \end{cases}
\end{align}

It is then trivial to show that at the limit of $L\rightarrow 0$, $mm'$ states and corresponding $N00N$ states approach minimal phase sensitivity $\delta \phi_{min}=1/\Delta m $ at optimal phase shifts
\begin{align}
   \phi _{op}=\begin{cases}                          
    \frac{(2n-1)}{2} \frac{\pi}{\Delta m}, & \text{if $\Delta m~ \text{is even}$};\\
    n\frac{\pi}{\Delta m}, & \text{if $\Delta m ~ \text{is odd}$}, ~n=1,2,\dots.
  \end{cases}
\end{align}

\begin{figure}
\centering
	\includegraphics[width=0.45\textwidth]{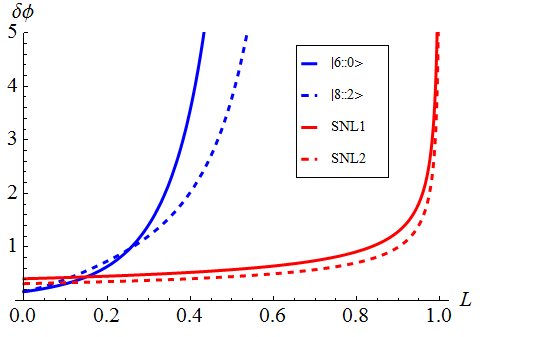}
	\caption{(Color online) Sensitivity for $mm'$ states as a function of loss rate $L$ in both arms of a two-mode interferometer. Solid blue curve corresponds to $\vert 6::0 \rangle$, dashed blue curve to $1/\sqrt{6 (1-L)}$, solid red line to $\vert 8::2 \rangle$ and dashed red curve to $1/\sqrt{10 (1-L)}$. Parity detection using $mm'$ states is worse than shot-noise under around 10\% loss. }
    \label{fig:sensitivity3}
\end{figure}

For a $mm'$ state or $N00N$ state to be able to beat the shot-noise limit under parity detection, we therefore should have
\begin{align}
\Delta m>\sqrt{m+m'}.
\end{align}
To meet the above criteria, here we choose $\Delta m=6$ and $\phi=\frac{\pi}{2 \Delta m}$, assuming $L_a=L_b$ for practical purposes. In Fig.~\ref{fig:sensitivity3} it can be seen that $\vert 6::0 \rangle$ and $\vert 8::2 \rangle$ give sub-shot-noise performances for loss less than about 10\%, and $\vert 6::0 \rangle$ gives higher sensitivity than $\vert 8::2 \rangle$ up to 25\% loss; for loss greater than 25\%,  $\vert 6::0 \rangle$ is outperformed by $\vert 8::2 \rangle$ but both are worse than shot-noise.

\subsection{Sensitivity for smaller loss}
\begin{figure}
\centering
	\includegraphics[width=0.45\textwidth]{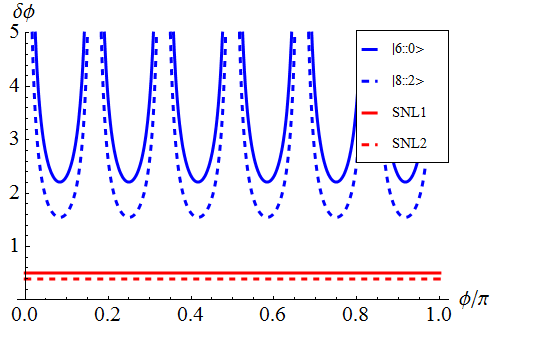}
	\caption{(Color online) Sensitivity of $mm'$ states as a function of phase shift $\phi$ from a two-mode interferometer under $35\%$ loss. Solid blue curve corresponds to $\vert 6::0 \rangle$, dashed blue curve to $1/\sqrt{6\times 0.65}$, solid red line to $\vert 8::2 \rangle$ and dashed red curve to $1/\sqrt{10\times 0.65}$. While the sensitivity of $mm'$ and $N00N$ states are worse than  respective shot-noise limit, $mm'$ is more robust as expected.}
    \label{fig:sensitivity1}
\end{figure}

Often we are more interested in low-loss regions where the sensitivity of $mm'$ states and $N00N$ states are comparable to the shot-noise limit. Fig.~\ref{fig:sensitivity1} shows that the sensitivity of $\vert 6::0 \rangle$ and $\vert 8::2 \rangle$ are noticeably worse than the respective shot-noise limit under moderate loss ($35\%$ in this case). Here $\vert 8::2 \rangle$ turns out to be more robust than $\vert 6::0 \rangle$ as predicted in Ref.~\cite{PhysRevA.78.063828}.

\begin{figure}
\centering
	\includegraphics[width=0.45\textwidth]{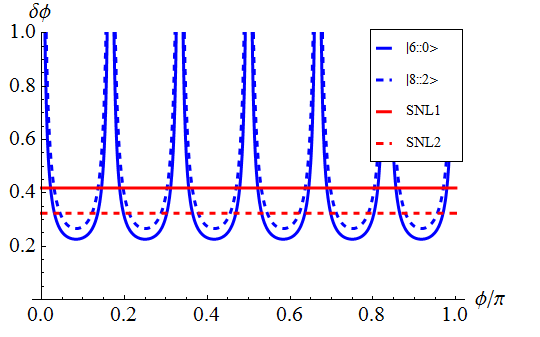}
	\caption{(Color online) Sensitivity of $mm'$ states as a function of phase shift $\phi$ from a two-mode interferometer under $5\%$ loss. Solid blue curve corresponds to $\vert 6::0 \rangle$, dashed blue curve to $1/\sqrt{6\times 0.95}$, solid red line to $\vert 8::2 \rangle$ and dashed red curve to $1/\sqrt{10\times 0.95}$. While the sensitivity of $mm'$ and $N00N$ states are better than their respective shot-noise limits, $N00N$ is more robust, which is unexpected.}
    \label{fig:sensitivity2}
	\end{figure}

This robustness, however, does not apply to situations where the loss is even smaller. In Fig.~\ref{fig:sensitivity2} we show the sensitivity of $\vert 6::0 \rangle$ and $\vert 8::2 \rangle$ under $5\%$ loss. Here both states give higher sensitivity than the shot-noise limit and $N00N$ is the best of all. In contrast, the result in Ref.~\cite{PhysRevA.78.063828} shows that, with a certain detection operator, $mm'$ states always give better sensitivity than using $N00N$ states and the shot-noise limit no matter how high the loss. This discrepancy indicates that the parity operator is not the optimal detector that always favors $mm'$ states (see conclusion for more discussion.) In addition, more states of the form $\vert m::m-6 \rangle$ with their highest sensitivity and corresponding shot-noise limit are shown in Table~\ref{tab:table1}.

\begin{table}
\caption{Optimal sensitivity of different $mm'$ states with $\Delta m=6$ and corresponding shot-noise limit $1/\sqrt{(m+m')(1-L)}$ under $5\%$ loss.}
\begin{ruledtabular}
\begin{tabular}{c c c c} 
$m$ & $m'$ & $\delta \phi$ & SNL \\ \hline
6 & 0 & 0.227 & 0.419 \\
8 & 2 & 0.266 & 0.324 \\
10 & 4 & 0.307 & 0.274 \\
12 & 6 & 0.348 & 0.242 \\
14 & 8 & 0.387 & 0.219 \\
\end{tabular}
\end{ruledtabular}

\label{tab:table1}
\end{table}

To conclude, $mm'$ states with large $m$ and small $m'$ or $N00N$ states under small loss perform with sub-shot-noise limit sensitivity.

\section{Conclusion}
We calculate the visibility and sensitivity of a phase-carrying signal by using $mm'$ states with parity detection in a lossy environment. In our calculation, we take the photon number difference between the two arms of the two-mode interferometer $\Delta m$ to be fixed to maintain the desired resolving power. Since visibility is not well defined for signal using parity detection, we use a visibility which is measured against signal without loss. To have high visibility, one should use $N00N$ states with large $N$ (when loss is low) and $mm'$ states with large $m+m'$ but small $\Delta m$ (for high loss). Considering only sensitivity, our calculation shows that $N00N$ states with large $N$ or $mm'$ states with large $\delta m$ under low loss are capable of performing with sub-shot-noise limit precision.

It is worthwhile to mention two points. First, modeling loss with a single fictitious beam splitter is sufficient for practical purposes. If one reverses the order of the phase shifter and the beam splitter in Fig.~\ref{fig:interferometer}, the same density matrix will be obtained as proved in Ref.~\cite{PhysRevLett.102.040403,PhysRevA.80.013825}. This also means our model is equivalent to a continuous loss model. Second, Ref. \cite{PhysRevA.78.063828} uses a detection operator that is carefully chosen so that it sums up all off-diagonal terms of the density matrix and provides sub-shot-noise sensitivity. Meanwhile the parity operator collects some of the diagonal and off-diagonal terms. The inclusion of diagonal terms may reduce the signal size and therefore visibility or sensitivity of phase information. However, such an operator in Ref.~\cite{PhysRevA.78.063828} is yet to be produced in a lab setting. On contrary, a lot of effort has been made to realize parity measurements. A straightforward parity measurement relies on high precision photon number-resolving detection at single-photon level, which has been demonstrated experimentally in near-infrared region \cite{LitaOpEx2008}. Alternative parity detection setups without number-resolving detectors have been proposed as well \cite{GerryPRA2005, PlickNJP2010}.

\section*{Acknowledgements}
We would like to acknowledge helpful discussions with K. L. Brown. This work is supported by the grant from NSF.


\bibliography{reference}

\begin{thebibliography}{29}%
\makeatletter
\providecommand \@ifxundefined [1]{%
 \@ifx{#1\undefined}
}%
\providecommand \@ifnum [1]{%
 \ifnum #1\expandafter \@firstoftwo
 \else \expandafter \@secondoftwo
 \fi
}%
\providecommand \@ifx [1]{%
 \ifx #1\expandafter \@firstoftwo
 \else \expandafter \@secondoftwo
 \fi
}%
\providecommand \natexlab [1]{#1}%
\providecommand \enquote  [1]{``#1''}%
\providecommand \bibnamefont  [1]{#1}%
\providecommand \bibfnamefont [1]{#1}%
\providecommand \citenamefont [1]{#1}%
\providecommand \href@noop [0]{\@secondoftwo}%
\providecommand \href [0]{\begingroup \@sanitize@url \@href}%
\providecommand \@href[1]{\@@startlink{#1}\@@href}%
\providecommand \@@href[1]{\endgroup#1\@@endlink}%
\providecommand \@sanitize@url [0]{\catcode `\\12\catcode `\$12\catcode
  `\&12\catcode `\#12\catcode `\^12\catcode `\_12\catcode `\%12\relax}%
\providecommand \@@startlink[1]{}%
\providecommand \@@endlink[0]{}%
\providecommand \url  [0]{\begingroup\@sanitize@url \@url }%
\providecommand \@url [1]{\endgroup\@href {#1}{\urlprefix }}%
\providecommand \urlprefix  [0]{URL }%
\providecommand \Eprint [0]{\href }%
\providecommand \doibase [0]{http://dx.doi.org/}%
\providecommand \selectlanguage [0]{\@gobble}%
\providecommand \bibinfo  [0]{\@secondoftwo}%
\providecommand \bibfield  [0]{\@secondoftwo}%
\providecommand \translation [1]{[#1]}%
\providecommand \BibitemOpen [0]{}%
\providecommand \bibitemStop [0]{}%
\providecommand \bibitemNoStop [0]{.\EOS\space}%
\providecommand \EOS [0]{\spacefactor3000\relax}%
\providecommand \BibitemShut  [1]{\csname bibitem#1\endcsname}%
\let\auto@bib@innerbib\@empty
\bibitem [{\citenamefont {Dowling}(2008)}]{Dowling2008}%
  \BibitemOpen
  \bibfield  {author} {\bibinfo {author} {\bibfnamefont {J.~P.}\ \bibnamefont
  {Dowling}},\ }\href {\doibase 10.1080/00107510802091298} {\bibfield
  {journal} {\bibinfo  {journal} {Contemp. Phys.}\ }\textbf {\bibinfo {volume}
  {49}},\ \bibinfo {pages} {125} (\bibinfo {year} {2008})}\BibitemShut
  {NoStop}%
\bibitem [{\citenamefont {Caves}(1982)}]{PhysRevD.26.1817}%
  \BibitemOpen
  \bibfield  {author} {\bibinfo {author} {\bibfnamefont {C.~M.}\ \bibnamefont
  {Caves}},\ }\href {\doibase 10.1103/PhysRevD.26.1817} {\bibfield  {journal}
  {\bibinfo  {journal} {Phys. Rev. D}\ }\textbf {\bibinfo {volume} {26}},\
  \bibinfo {pages} {1817} (\bibinfo {year} {1982})}\BibitemShut {NoStop}%
\bibitem [{\citenamefont {Boto}\ \emph {et~al.}(2000)\citenamefont {Boto},
  \citenamefont {Kok}, \citenamefont {Abrams}, \citenamefont {Braunstein},
  \citenamefont {Williams},\ and\ \citenamefont
  {Dowling}}]{PhysRevLett.85.2733}%
  \BibitemOpen
  \bibfield  {author} {\bibinfo {author} {\bibfnamefont {A.~N.}\ \bibnamefont
  {Boto}}, \bibinfo {author} {\bibfnamefont {P.}~\bibnamefont {Kok}}, \bibinfo
  {author} {\bibfnamefont {D.~S.}\ \bibnamefont {Abrams}}, \bibinfo {author}
  {\bibfnamefont {S.~L.}\ \bibnamefont {Braunstein}}, \bibinfo {author}
  {\bibfnamefont {C.~P.}\ \bibnamefont {Williams}}, \ and\ \bibinfo {author}
  {\bibfnamefont {J.~P.}\ \bibnamefont {Dowling}},\ }\href {\doibase
  10.1103/PhysRevLett.85.2733} {\bibfield  {journal} {\bibinfo  {journal}
  {Phys. Rev. Lett.}\ }\textbf {\bibinfo {volume} {85}},\ \bibinfo {pages}
  {2733} (\bibinfo {year} {2000})}\BibitemShut {NoStop}%
\bibitem [{\citenamefont {Durkin}\ and\ \citenamefont
  {Dowling}(2007)}]{PhysRevLett.99.070801}%
  \BibitemOpen
  \bibfield  {author} {\bibinfo {author} {\bibfnamefont {G.~A.}\ \bibnamefont
  {Durkin}}\ and\ \bibinfo {author} {\bibfnamefont {J.~P.}\ \bibnamefont
  {Dowling}},\ }\href {\doibase 10.1103/PhysRevLett.99.070801} {\bibfield
  {journal} {\bibinfo  {journal} {Phys. Rev. Lett.}\ }\textbf {\bibinfo
  {volume} {99}},\ \bibinfo {pages} {070801} (\bibinfo {year}
  {2007})}\BibitemShut {NoStop}%
\bibitem [{\citenamefont {Rubin}\ and\ \citenamefont
  {Kaushik}(2007)}]{PhysRevA.75.053805}%
  \BibitemOpen
  \bibfield  {author} {\bibinfo {author} {\bibfnamefont {M.~A.}\ \bibnamefont
  {Rubin}}\ and\ \bibinfo {author} {\bibfnamefont {S.}~\bibnamefont
  {Kaushik}},\ }\href {\doibase 10.1103/PhysRevA.75.053805} {\bibfield
  {journal} {\bibinfo  {journal} {Phys. Rev. A}\ }\textbf {\bibinfo {volume}
  {75}},\ \bibinfo {pages} {053805} (\bibinfo {year} {2007})}\BibitemShut
  {NoStop}%
\bibitem [{\citenamefont {Gilbert}\ \emph {et~al.}(2007)\citenamefont
  {Gilbert}, \citenamefont {Hamrick},\ and\ \citenamefont
  {Weinstein}}]{Gilbert2007}%
  \BibitemOpen
  \bibfield  {author} {\bibinfo {author} {\bibfnamefont {G.}~\bibnamefont
  {Gilbert}}, \bibinfo {author} {\bibfnamefont {M.}~\bibnamefont {Hamrick}}, \
  and\ \bibinfo {author} {\bibfnamefont {Y.~S.}\ \bibnamefont {Weinstein}},\
  }\href@noop {} {\bibfield  {journal} {\bibinfo  {journal} {Proc. SPIE}\
  }\textbf {\bibinfo {volume} {6573}},\ \bibinfo {pages} {6573K} (\bibinfo
  {year} {2007})}\BibitemShut {NoStop}%
\bibitem [{\citenamefont {Parks}\ \emph {et~al.}(2007)\citenamefont {Parks},
  \citenamefont {Spence}, \citenamefont {Troupe},\ and\ \citenamefont
  {Rodecap}}]{RevSciInstrum.76.043103}%
  \BibitemOpen
  \bibfield  {author} {\bibinfo {author} {\bibfnamefont {A.~D.}\ \bibnamefont
  {Parks}}, \bibinfo {author} {\bibfnamefont {S.~E.}\ \bibnamefont {Spence}},
  \bibinfo {author} {\bibfnamefont {J.~E.}\ \bibnamefont {Troupe}}, \ and\
  \bibinfo {author} {\bibfnamefont {N.~J.}\ \bibnamefont {Rodecap}},\ }\href
  {\doibase 10.1063/1.1879332} {\bibfield  {journal} {\bibinfo  {journal} {Rev.
  Sci. Instrum}\ }\textbf {\bibinfo {volume} {76}},\ \bibinfo {pages} {043103}
  (\bibinfo {year} {2007})}\BibitemShut {NoStop}%
\bibitem [{\citenamefont {Huver}\ \emph {et~al.}(2008)\citenamefont {Huver},
  \citenamefont {Wildfeuer},\ and\ \citenamefont
  {Dowling}}]{PhysRevA.78.063828}%
  \BibitemOpen
  \bibfield  {author} {\bibinfo {author} {\bibfnamefont {S.~D.}\ \bibnamefont
  {Huver}}, \bibinfo {author} {\bibfnamefont {C.~F.}\ \bibnamefont
  {Wildfeuer}}, \ and\ \bibinfo {author} {\bibfnamefont {J.~P.}\ \bibnamefont
  {Dowling}},\ }\href {\doibase 10.1103/PhysRevA.78.063828} {\bibfield
  {journal} {\bibinfo  {journal} {Phys. Rev. A}\ }\textbf {\bibinfo {volume}
  {78}},\ \bibinfo {pages} {063828} (\bibinfo {year} {2008})}\BibitemShut
  {NoStop}%
\bibitem [{\citenamefont {Resch}\ \emph {et~al.}(2007)\citenamefont {Resch},
  \citenamefont {Pregnell}, \citenamefont {Prevedel}, \citenamefont
  {Gilchrist}, \citenamefont {Pryde}, \citenamefont {O'Brien},\ and\
  \citenamefont {White}}]{ReschPRL2007}%
  \BibitemOpen
  \bibfield  {author} {\bibinfo {author} {\bibfnamefont {K.~J.}\ \bibnamefont
  {Resch}}, \bibinfo {author} {\bibfnamefont {K.~L.}\ \bibnamefont {Pregnell}},
  \bibinfo {author} {\bibfnamefont {R.}~\bibnamefont {Prevedel}}, \bibinfo
  {author} {\bibfnamefont {A.}~\bibnamefont {Gilchrist}}, \bibinfo {author}
  {\bibfnamefont {G.~J.}\ \bibnamefont {Pryde}}, \bibinfo {author}
  {\bibfnamefont {J.~L.}\ \bibnamefont {O'Brien}}, \ and\ \bibinfo {author}
  {\bibfnamefont {A.~G.}\ \bibnamefont {White}},\ }\href {\doibase
  10.1103/PhysRevLett.98.223601} {\bibfield  {journal} {\bibinfo  {journal}
  {Phys. Rev. Lett.}\ }\textbf {\bibinfo {volume} {98}},\ \bibinfo {pages}
  {223601} (\bibinfo {year} {2007})}\BibitemShut {NoStop}%
\bibitem [{\citenamefont {Giovannetti}\ \emph {et~al.}(2009)\citenamefont
  {Giovannetti}, \citenamefont {Lloyd}, \citenamefont {Maccone},\ and\
  \citenamefont {Shapiro}}]{GiovannettiPRA2009}%
  \BibitemOpen
  \bibfield  {author} {\bibinfo {author} {\bibfnamefont {V.}~\bibnamefont
  {Giovannetti}}, \bibinfo {author} {\bibfnamefont {S.}~\bibnamefont {Lloyd}},
  \bibinfo {author} {\bibfnamefont {L.}~\bibnamefont {Maccone}}, \ and\
  \bibinfo {author} {\bibfnamefont {J.~H.}\ \bibnamefont {Shapiro}},\ }\href
  {\doibase 10.1103/PhysRevA.79.013827} {\bibfield  {journal} {\bibinfo
  {journal} {Phys. Rev. A}\ }\textbf {\bibinfo {volume} {79}},\ \bibinfo
  {pages} {013827} (\bibinfo {year} {2009})}\BibitemShut {NoStop}%
\bibitem [{\citenamefont {Nagata}\ \emph {et~al.}(2007)\citenamefont {Nagata},
  \citenamefont {Okamoto}, \citenamefont {O'Brien}, \citenamefont {Sasaki},\
  and\ \citenamefont {Takeuchi}}]{NagataSci2007}%
  \BibitemOpen
  \bibfield  {author} {\bibinfo {author} {\bibfnamefont {T.}~\bibnamefont
  {Nagata}}, \bibinfo {author} {\bibfnamefont {R.}~\bibnamefont {Okamoto}},
  \bibinfo {author} {\bibfnamefont {J.~L.}\ \bibnamefont {O'Brien}}, \bibinfo
  {author} {\bibfnamefont {K.}~\bibnamefont {Sasaki}}, \ and\ \bibinfo {author}
  {\bibfnamefont {S.}~\bibnamefont {Takeuchi}},\ }\href {\doibase
  10.1126/science.1138007} {\bibfield  {journal} {\bibinfo  {journal}
  {Science}\ }\textbf {\bibinfo {volume} {316}},\ \bibinfo {pages} {726}
  (\bibinfo {year} {2007})}\BibitemShut {NoStop}%
\bibitem [{\citenamefont {Gerry}(2000)}]{PhysRevA.61.043811}%
  \BibitemOpen
  \bibfield  {author} {\bibinfo {author} {\bibfnamefont {C.~C.}\ \bibnamefont
  {Gerry}},\ }\href {\doibase 10.1103/PhysRevA.61.043811} {\bibfield  {journal}
  {\bibinfo  {journal} {Phys. Rev. A}\ }\textbf {\bibinfo {volume} {61}},\
  \bibinfo {pages} {043811} (\bibinfo {year} {2000})}\BibitemShut {NoStop}%
\bibitem [{\citenamefont {Campos}\ \emph {et~al.}(2003)\citenamefont {Campos},
  \citenamefont {Gerry},\ and\ \citenamefont {Benmoussa}}]{PhysRevA.68.023810}%
  \BibitemOpen
  \bibfield  {author} {\bibinfo {author} {\bibfnamefont {R.~A.}\ \bibnamefont
  {Campos}}, \bibinfo {author} {\bibfnamefont {C.~C.}\ \bibnamefont {Gerry}}, \
  and\ \bibinfo {author} {\bibfnamefont {A.}~\bibnamefont {Benmoussa}},\ }\href
  {\doibase 10.1103/PhysRevA.68.023810} {\bibfield  {journal} {\bibinfo
  {journal} {Phys. Rev. A}\ }\textbf {\bibinfo {volume} {68}},\ \bibinfo
  {pages} {023810} (\bibinfo {year} {2003})}\BibitemShut {NoStop}%
\bibitem [{\citenamefont {Gerry}\ and\ \citenamefont
  {Mimih}(2010)}]{doi:10.1080/00107514.2010.509995}%
  \BibitemOpen
  \bibfield  {author} {\bibinfo {author} {\bibfnamefont {C.~C.}\ \bibnamefont
  {Gerry}}\ and\ \bibinfo {author} {\bibfnamefont {J.}~\bibnamefont {Mimih}},\
  }\href {\doibase 10.1080/00107514.2010.509995} {\bibfield  {journal}
  {\bibinfo  {journal} {Contemporary Physics}\ }\textbf {\bibinfo {volume}
  {51}},\ \bibinfo {pages} {497} (\bibinfo {year} {2010})}\BibitemShut
  {NoStop}%
\bibitem [{\citenamefont {Yurke}\ \emph {et~al.}(1986)\citenamefont {Yurke},
  \citenamefont {McCall},\ and\ \citenamefont {Klauder}}]{PhysRevA.33.4033}%
  \BibitemOpen
  \bibfield  {author} {\bibinfo {author} {\bibfnamefont {B.}~\bibnamefont
  {Yurke}}, \bibinfo {author} {\bibfnamefont {S.~L.}\ \bibnamefont {McCall}}, \
  and\ \bibinfo {author} {\bibfnamefont {J.~R.}\ \bibnamefont {Klauder}},\
  }\href {\doibase 10.1103/PhysRevA.33.4033} {\bibfield  {journal} {\bibinfo
  {journal} {Phys. Rev. A}\ }\textbf {\bibinfo {volume} {33}},\ \bibinfo
  {pages} {4033} (\bibinfo {year} {1986})}\BibitemShut {NoStop}%
\bibitem [{\citenamefont {Kim}\ \emph {et~al.}(1998)\citenamefont {Kim},
  \citenamefont {Pfister}, \citenamefont {Holland}, \citenamefont {Noh},\ and\
  \citenamefont {Hall}}]{PhysRevA.57.4004}%
  \BibitemOpen
  \bibfield  {author} {\bibinfo {author} {\bibfnamefont {T.}~\bibnamefont
  {Kim}}, \bibinfo {author} {\bibfnamefont {O.}~\bibnamefont {Pfister}},
  \bibinfo {author} {\bibfnamefont {M.~J.}\ \bibnamefont {Holland}}, \bibinfo
  {author} {\bibfnamefont {J.}~\bibnamefont {Noh}}, \ and\ \bibinfo {author}
  {\bibfnamefont {J.~L.}\ \bibnamefont {Hall}},\ }\href {\doibase
  10.1103/PhysRevA.57.4004} {\bibfield  {journal} {\bibinfo  {journal} {Phys.
  Rev. A}\ }\textbf {\bibinfo {volume} {57}},\ \bibinfo {pages} {4004}
  (\bibinfo {year} {1998})}\BibitemShut {NoStop}%
\bibitem [{\citenamefont {Chiruvelli}\ and\ \citenamefont
  {Lee}(2011)}]{doi:10.1080/09500340.2011.585251}%
  \BibitemOpen
  \bibfield  {author} {\bibinfo {author} {\bibfnamefont {A.}~\bibnamefont
  {Chiruvelli}}\ and\ \bibinfo {author} {\bibfnamefont {H.}~\bibnamefont
  {Lee}},\ }\href {\doibase 10.1080/09500340.2011.585251} {\bibfield  {journal}
  {\bibinfo  {journal} {Journal of Modern Optics}\ }\textbf {\bibinfo {volume}
  {58}},\ \bibinfo {pages} {945} (\bibinfo {year} {2011})}\BibitemShut
  {NoStop}%
\bibitem [{\citenamefont {Bevington}(1969)}]{Bevington1969}%
  \BibitemOpen
  \bibfield  {author} {\bibinfo {author} {\bibfnamefont {P.~R.}\ \bibnamefont
  {Bevington}},\ }\href@noop {} {\emph {\bibinfo {title} {Data Reduction and
  Error Analysis for the Physical Sciences}}}\ (\bibinfo  {publisher}
  {McGraw-Hill, New York},\ \bibinfo {year} {1969})\BibitemShut {NoStop}%
\bibitem [{\citenamefont {Loudon}(2000)}]{Loudon2000}%
  \BibitemOpen
  \bibfield  {author} {\bibinfo {author} {\bibfnamefont {R.}~\bibnamefont
  {Loudon}},\ }\href@noop {} {\emph {\bibinfo {title} {The Quantum Theory of
  Light}}},\ \bibinfo {edition} {third edition}\ ed.\ (\bibinfo  {publisher}
  {Oxford University Press},\ \bibinfo {year} {2000})\BibitemShut {NoStop}%
\bibitem [{\citenamefont {Gardiner}\ and\ \citenamefont
  {Zoller}(2004)}]{Gardiner2004}%
  \BibitemOpen
  \bibfield  {author} {\bibinfo {author} {\bibfnamefont {C.~W.}\ \bibnamefont
  {Gardiner}}\ and\ \bibinfo {author} {\bibfnamefont {P.}~\bibnamefont
  {Zoller}},\ }\href@noop {} {\emph {\bibinfo {title} {Quantum Noise}}}\
  (\bibinfo  {publisher} {Springer, Berilin},\ \bibinfo {year}
  {2004})\BibitemShut {NoStop}%
\bibitem [{\citenamefont {Gerry}\ and\ \citenamefont
  {Knight}(2005)}]{Gerry2005}%
  \BibitemOpen
  \bibfield  {author} {\bibinfo {author} {\bibfnamefont {C.~C.}\ \bibnamefont
  {Gerry}}\ and\ \bibinfo {author} {\bibfnamefont {P.}~\bibnamefont {Knight}},\
  }\href@noop {} {\emph {\bibinfo {title} {Introductory quantum optics}}}\
  (\bibinfo  {publisher} {Cambridge University Press},\ \bibinfo {year}
  {2005})\BibitemShut {NoStop}%
\bibitem [{\citenamefont {Lee}\ \emph {et~al.}(2009)\citenamefont {Lee},
  \citenamefont {Huver}, \citenamefont {Lee}, \citenamefont {Kaplan},
  \citenamefont {McCracken}, \citenamefont {Min}, \citenamefont {Uskov},
  \citenamefont {Wildfeuer}, \citenamefont {Veronis},\ and\ \citenamefont
  {Dowling}}]{LeePRA2009}%
  \BibitemOpen
  \bibfield  {author} {\bibinfo {author} {\bibfnamefont {T.-W.}\ \bibnamefont
  {Lee}}, \bibinfo {author} {\bibfnamefont {S.~D.}\ \bibnamefont {Huver}},
  \bibinfo {author} {\bibfnamefont {H.}~\bibnamefont {Lee}}, \bibinfo {author}
  {\bibfnamefont {L.}~\bibnamefont {Kaplan}}, \bibinfo {author} {\bibfnamefont
  {S.~B.}\ \bibnamefont {McCracken}}, \bibinfo {author} {\bibfnamefont
  {C.}~\bibnamefont {Min}}, \bibinfo {author} {\bibfnamefont {D.~B.}\
  \bibnamefont {Uskov}}, \bibinfo {author} {\bibfnamefont {C.~F.}\ \bibnamefont
  {Wildfeuer}}, \bibinfo {author} {\bibfnamefont {G.}~\bibnamefont {Veronis}},
  \ and\ \bibinfo {author} {\bibfnamefont {J.~P.}\ \bibnamefont {Dowling}},\
  }\href {\doibase 10.1103/PhysRevA.80.063803} {\bibfield  {journal} {\bibinfo
  {journal} {Phys. Rev. A}\ }\textbf {\bibinfo {volume} {80}},\ \bibinfo
  {pages} {063803} (\bibinfo {year} {2009})}\BibitemShut {NoStop}%
\bibitem [{\citenamefont {Dorner}\ \emph {et~al.}(2009)\citenamefont {Dorner},
  \citenamefont {Demkowicz-Dobrzanski}, \citenamefont {Smith}, \citenamefont
  {Lundeen}, \citenamefont {Wasilewski}, \citenamefont {Banaszek},\ and\
  \citenamefont {Walmsley}}]{PhysRevLett.102.040403}%
  \BibitemOpen
  \bibfield  {author} {\bibinfo {author} {\bibfnamefont {U.}~\bibnamefont
  {Dorner}}, \bibinfo {author} {\bibfnamefont {R.}~\bibnamefont
  {Demkowicz-Dobrzanski}}, \bibinfo {author} {\bibfnamefont {B.~J.}\
  \bibnamefont {Smith}}, \bibinfo {author} {\bibfnamefont {J.~S.}\ \bibnamefont
  {Lundeen}}, \bibinfo {author} {\bibfnamefont {W.}~\bibnamefont {Wasilewski}},
  \bibinfo {author} {\bibfnamefont {K.}~\bibnamefont {Banaszek}}, \ and\
  \bibinfo {author} {\bibfnamefont {I.~A.}\ \bibnamefont {Walmsley}},\ }\href
  {\doibase 10.1103/PhysRevLett.102.040403} {\bibfield  {journal} {\bibinfo
  {journal} {Phys. Rev. Lett.}\ }\textbf {\bibinfo {volume} {102}},\ \bibinfo
  {pages} {040403} (\bibinfo {year} {2009})}\BibitemShut {NoStop}%
\bibitem [{\citenamefont {Demkowicz-Dobrzanski}\ \emph
  {et~al.}(2009)\citenamefont {Demkowicz-Dobrzanski}, \citenamefont {Dorner},
  \citenamefont {Smith}, \citenamefont {Lundeen}, \citenamefont {Wasilewski},
  \citenamefont {Banaszek},\ and\ \citenamefont
  {Walmsley}}]{PhysRevA.80.013825}%
  \BibitemOpen
  \bibfield  {author} {\bibinfo {author} {\bibfnamefont {R.}~\bibnamefont
  {Demkowicz-Dobrzanski}}, \bibinfo {author} {\bibfnamefont {U.}~\bibnamefont
  {Dorner}}, \bibinfo {author} {\bibfnamefont {B.~J.}\ \bibnamefont {Smith}},
  \bibinfo {author} {\bibfnamefont {J.~S.}\ \bibnamefont {Lundeen}}, \bibinfo
  {author} {\bibfnamefont {W.}~\bibnamefont {Wasilewski}}, \bibinfo {author}
  {\bibfnamefont {K.}~\bibnamefont {Banaszek}}, \ and\ \bibinfo {author}
  {\bibfnamefont {I.~A.}\ \bibnamefont {Walmsley}},\ }\href {\doibase
  10.1103/PhysRevA.80.013825} {\bibfield  {journal} {\bibinfo  {journal} {Phys.
  Rev. A}\ }\textbf {\bibinfo {volume} {80}},\ \bibinfo {pages} {013825}
  (\bibinfo {year} {2009})}\BibitemShut {NoStop}%
\bibitem [{\citenamefont {Sakurai}(1993)}]{Sakurai1993}%
  \BibitemOpen
  \bibfield  {author} {\bibinfo {author} {\bibfnamefont {J.~J.}\ \bibnamefont
  {Sakurai}},\ }\href@noop {} {\emph {\bibinfo {title} {Modern Quantum
  Mechanics}}}\ (\bibinfo  {publisher} {Addison Wesley},\ \bibinfo {year}
  {1993})\BibitemShut {NoStop}%
\bibitem [{\citenamefont {Gradshteyn}\ and\ \citenamefont
  {Ryzhik}(1965)}]{Gradshteyn1965}%
  \BibitemOpen
  \bibfield  {author} {\bibinfo {author} {\bibfnamefont {I.~S.}\ \bibnamefont
  {Gradshteyn}}\ and\ \bibinfo {author} {\bibfnamefont {I.~M.}\ \bibnamefont
  {Ryzhik}},\ }\href@noop {} {\emph {\bibinfo {title} {Table of Integrals
  Series and Products}}},\ \bibinfo {edition} {4th}\ ed.\ (\bibinfo
  {publisher} {Academic press},\ \bibinfo {year} {1965})\BibitemShut {NoStop}%
\bibitem [{\citenamefont {Lita}\ \emph {et~al.}(2008)\citenamefont {Lita},
  \citenamefont {Miller},\ and\ \citenamefont {Nam}}]{LitaOpEx2008}%
  \BibitemOpen
  \bibfield  {author} {\bibinfo {author} {\bibfnamefont {A.~E.}\ \bibnamefont
  {Lita}}, \bibinfo {author} {\bibfnamefont {A.~J.}\ \bibnamefont {Miller}}, \
  and\ \bibinfo {author} {\bibfnamefont {S.~W.}\ \bibnamefont {Nam}},\
  }\href@noop {} {\bibfield  {journal} {\bibinfo  {journal} {Optics Express}\
  }\textbf {\bibinfo {volume} {16}},\ \bibinfo {pages} {3032} (\bibinfo {year}
  {2008})}\BibitemShut {NoStop}%
\bibitem [{\citenamefont {Gerry}\ \emph {et~al.}(2005)\citenamefont {Gerry},
  \citenamefont {Benmoussa},\ and\ \citenamefont {Campos}}]{GerryPRA2005}%
  \BibitemOpen
  \bibfield  {author} {\bibinfo {author} {\bibfnamefont {C.~C.}\ \bibnamefont
  {Gerry}}, \bibinfo {author} {\bibfnamefont {A.}~\bibnamefont {Benmoussa}}, \
  and\ \bibinfo {author} {\bibfnamefont {R.~A.}\ \bibnamefont {Campos}},\
  }\href {\doibase 10.1103/PhysRevA.72.053818} {\bibfield  {journal} {\bibinfo
  {journal} {Phys. Rev. A}\ }\textbf {\bibinfo {volume} {72}},\ \bibinfo
  {pages} {053818} (\bibinfo {year} {2005})}\BibitemShut {NoStop}%
\bibitem [{\citenamefont {Plick}\ \emph {et~al.}(2010)\citenamefont {Plick},
  \citenamefont {Anisimov}, \citenamefont {Dowling}, \citenamefont {Lee},\ and\
  \citenamefont {Agarwal}}]{PlickNJP2010}%
  \BibitemOpen
  \bibfield  {author} {\bibinfo {author} {\bibfnamefont {W.~N.}\ \bibnamefont
  {Plick}}, \bibinfo {author} {\bibfnamefont {P.~M.}\ \bibnamefont {Anisimov}},
  \bibinfo {author} {\bibfnamefont {J.~P.}\ \bibnamefont {Dowling}}, \bibinfo
  {author} {\bibfnamefont {H.}~\bibnamefont {Lee}}, \ and\ \bibinfo {author}
  {\bibfnamefont {G.~S.}\ \bibnamefont {Agarwal}},\ }\href@noop {} {\bibfield
  {journal} {\bibinfo  {journal} {New Journal of Physics}\ }\textbf {\bibinfo
  {volume} {12}},\ \bibinfo {pages} {113025} (\bibinfo {year}
  {2010})}\BibitemShut {NoStop}%
\end{thebibliography}%
\end{document}